\documentclass[aps,10pt,prl,reprint,twocolumn,showpacs,floatfix,superscriptaddress,longbibliography]{revtex4-2}
\usepackage{amsfonts,amsmath,amssymb}

\usepackage[margin=20mm,inner=20mm,marginpar=10mm]{geometry}
\usepackage{graphicx, epstopdf}
\usepackage{microtype}
\usepackage[figuresright]{rotating}
\usepackage{setspace}
\usepackage{textcomp}
\usepackage{times}
\usepackage[normalem]{ulem}
\usepackage[abs]{overpic}
\usepackage{color}

\usepackage[T1]{fontenc}
\usepackage[utf8]{inputenc}
\usepackage{chemformula}
\usepackage{soul}
\usepackage{siunitx}
\usepackage{bm}
\usepackage{booktabs}
\usepackage{multirow}
\usepackage[version=4]{mhchem}

\newlength{\figwidth}
\begin{document}

\title{Diffracting molecular matter-waves at deep-ultraviolet standing-light waves}

\author{Ksenija Simonovi\'c}
\email[]{ksenija.simonovic@univie.ac.at}
\affiliation{University of Vienna, Faculty of Physics, VDS, VCQ, Boltzmanngasse 5, A-1090 Vienna, Austria}

\author{Richard Ferstl}
\affiliation{University of Vienna, Faculty of Physics, VDS, VCQ, Boltzmanngasse 5, A-1090 Vienna, Austria}

\author{Alfredo Di Silvestro}
\affiliation{Department of Chemistry, University of Basel, St. Johannsring 19, 4056 Basel, Switzerland}

\author{Marcel Mayor}
\affiliation{Department of Chemistry, University of Basel, St. Johannsring 19, 4056 Basel, Switzerland}

\author{Lukas Martinetz}
\affiliation{University of Duisburg-Essen, Lotharstraße 1, 47048 Duisburg, Germany}

\author{Klaus Hornberger}
\affiliation{University of Duisburg-Essen, Lotharstraße 1, 47048 Duisburg, Germany}

\author{Benjamin A. Stickler}
\affiliation{Ulm University, Institute for Complex Quantum Systems and Center for Integrated Quantum Science and Technology, Albert-Einstein-Allee 11, 89069 Ulm, Germany}

\author{Christian Brand}
\affiliation{German Aerospace Center (DLR), Institute of Quantum Technologies, Wilhelm-Runge-Straße 10, 89081 Ulm, Germany}

\author{Markus Arndt}
\email[]{markus.arndt@univie.ac.at}
\affiliation{University of Vienna, Faculty of Physics, VDS, VCQ, Boltzmanngasse 5, A-1090 Vienna, Austria}%

\date{\today}

\begin{abstract}   
Matter-wave interferometry with molecules is intriguing both because it demonstrates a fundamental quantum phenomenon and because it opens avenues to quantum-enhanced measurements in physical chemistry. One great challenge in such experiments is to establish matter-wave beam splitting mechanisms that are efficient and applicable to a wide range of particles. In the past, continuous standing light waves in the visible spectral range were used predominantly as phase gratings, while pulsed vacuum ultraviolet light found applications in photo-ionisation gratings. Here, we explore the regime of continuous, intense deep-ultraviolet (>\SI{1}{MW\per cm^2}, \SI{266}{nm}) light masks, where a rich variety of photo-physical and photo-chemical phenomena and relaxation pathways must be considered. The improved understanding of the mechanisms in this interaction opens new potential pathways to protein interferometry and to matter-wave enhanced sensing of molecular properties.
\end{abstract}

\maketitle

\section{Introduction}
Shortly after Louis de Broglie's prediction that one needs to 
'associate a periodical phenomenon with any isolated portion of matter or energy' and 
that we should see this 'in phase with a wave'
\cite{de_broglie_waves_1923}, matter-waves were experimentally confirmed for electrons \cite{davisson_diffraction_1927,Thomson1927},  neutral He atoms and \ch{H2} molecules~\cite{estermann_beugung_1930}, as well as for neutrons\cite{Halban1936}.
De Broglie's revolutionary proposal about the wave behaviour of matter~\cite{de_broglie_waves_1923} started the theoretical formulation of modern quantum physics~\cite{schrodinger_undulatory_1926}
and quantum chemistry, where this idea is at the heart of molecular bond and orbital theory~\cite{mulliken_assignment_1928,pauling_nature_1931}.
While in chemistry electron quantum waves are usually confined inside an atom or molecule, a whole research field has evolved around the question how to describe the center-of-mass motion of single and composite systems -- from electrons~\cite{hasselbach_progress_2009} over neutrons~\cite{rauch_neutron_2015} and atoms~\cite{cronin_optics_2009,tino_atom_2014} to complex molecules~\cite{hornberger_colloquium_2012} or even antimatter~\cite{sala_first_2019}.  

Here, we are focusing on new tools for the quantum coherent manipulation of the centre-of-mass motion of large molecules, inspired by advances in atom interferometry and progress in the diffraction of cold dimers~\cite{chapman_optics_1995}, small noble gas clusters~\cite{schollkopf_nondestructive_1994,Zhao2013}, and large molecules~\cite{arndt_waveparticle_1999}.
Numerous molecule interferometers have been built throughout the last two decades to explore molecular transition strengths~\cite{lisdat_realization_2000,liu_ramsey-borde_2010}, to study the quantum wave nature of fullerenes~\cite{brezger_matter-wave_2002}, vitamins~\cite{mairhofer_quantum-assisted_2017}, polypeptides~\cite{shayeghi_matter-wave_2020}, clusters of organic molecules~\cite{haslinger_universal_2013} or tailor-made compounds with masses even beyond 25\,kDa~\cite{fein_quantum_2019}.
\begin{figure*}[th]
    \centering
    \includegraphics[width=0.9\linewidth]{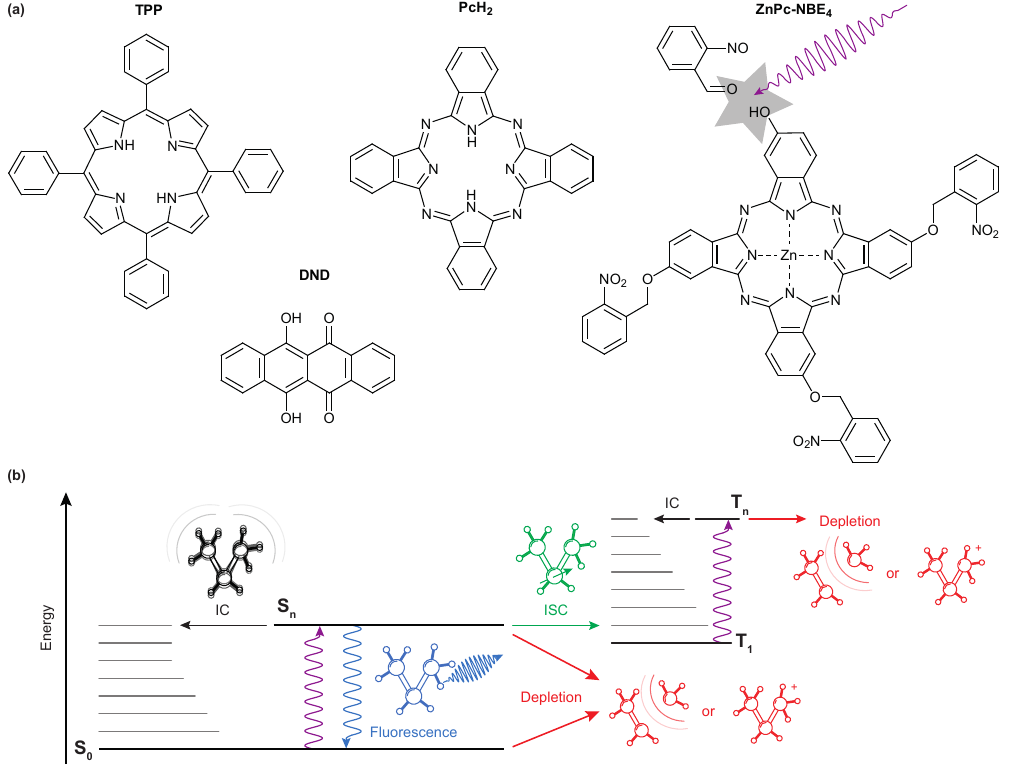}
    \caption{\textbf{(a)}: Molecules explored in this experiment, from left to right: \emph{meso}-Tetraphenylporphyrin (\ch{TPP}), 6,11-Dihydroxy-5,12-naphthacenedione (\ch{DND}), metal-free phthalocyanine (\ch{PcH2}), a zinc-coordinated phthalocyanine derivative (\ch{ZnPc-NBE4}). \textbf{(b)}: Possible internal relaxation pathways after deep-ultraviolet photo-excitation. After electronic excitation, the emission of a fluorescence photon adds a randomly oriented photon recoil to the molecule, blurring the respective diffraction peaks. This is not the case for non-radiative processes, such as internal conversion and inter-system crossing. Fragmentation or ionisation may occur from any excited state or a hot ground state molecule, removing it effectively from the beam.}
    \label{fig1-molecules_processes}
\end{figure*}
A variety of recent experiments in physical chemistry have focused on the analysis of molecules and clusters in classical and quantum beam deflectometry\cite{dugourd_beam_2001,compagnon_permanent_2001,antoine_electric_2002,shayeghi_nature_2015,fuchs_influence_2021,rivic_discriminating_2021}. These studies find a valuable complement in matter-wave interferometry which also allows measuring the electric~\cite{gring_influence_2010}, magnetic~\cite{fein_nanoscale_2022}, optical~\cite{eibenberger_absolute_2014} or structural properties~\cite{gerlich_matter-wave_2008,gring_influence_2010, tuxen_quantum_2010} of complex molecules via deflection of fine-grained quantum interference fringes.
  
Extending matter-wave interferometry to an even larger set of molecules requires new methods for molecular beam generation, beam splitters, and efficient single-molecule detectors. Here, we focus on how to realise deep ultraviolet beam splitters and how they interact with the rich set of internal molecular properties. Inspired by early achievements in atom optics~\cite{keith_diffraction_1988, carnal_youngs_1991}, nanomechanical masks were already successfully used to manipulate molecular beams ~\cite{arndt_waveparticle_1999,reisinger_poissons_2009,brand_atomically_2015, luski_vortex_2021}. While these nanostructures are very well suited for many atoms and molecules with low electric polarisability and dipole moments~\cite{knobloch_role_2017,simonovic_diffraction_2024}, optical gratings cannot be clogged or destroyed. They are perfectly periodic, adjustable in situ and in real time and they may also exploit internal states that nanomasks would not be sensitive to. 

Inspired by prior experiments in atom optics ~\cite{moskowitz_diffraction_1983, gould_diffraction_1986, rasel_atom_1995} and electron  optics\cite{freimund_observation_2001}, optical phase gratings were realised for molecular beams of fullerenes ~\cite{nairz_diffraction_2001} and even antibiotics~\cite{brand_bragg_2020} and pulsed vacuum-ultraviolet photo-ionisation gratings as matter-wave beam splitters for organic clusters\cite{haslinger_universal_2013} and polypeptides\cite{shayeghi_matter-wave_2020}. 
\begin{figure*}[tbh]
    \centering
    \includegraphics[width=0.7\linewidth]{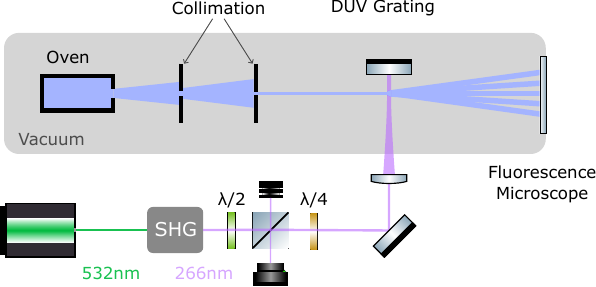}
    \caption{Setup of the experiment: A thermal molecular beam is collimated to a divergence of \SI{5}{\micro \radian} to approximate a plane-parallel matter-wave. The molecules are diffracted at a deep ultraviolet grating which is generated as a standing light wave of a high-power continuous frequency doubled laser. The diffracted molecules generate a mass density pattern on the window of that vacuum chamber, which is imaged using fluorescence microscopy. During diffraction, the matter-wave beam splitter imparts a transverse momentum of $\Delta p =\pm n \hbar k_L$, with the integer $n$ depending on the details of the process.}
    \centering
    \label{fig2-setup}
\end{figure*}
Here, we study the regime of continuous, high-intensity deep-ultraviolet (DUV) light masks.
The wavelength of $\lambda_{\rm L} = \SI{266}{nm}$ is close to a (very broad) electronic transition in many aromatic molecules and high-power laser light can be generated with high coherence and in a good beam profile by second harmonic generation of a diode pumped solid state laser. 
For thermal beams of molecules with an absorption cross-section around $\sigma_\mathrm{abs}\simeq 10^{-16}\,\mathrm{cm}^2$ and velocities in the range of 100-300\,m/s, laser intensities around \unit{1 MW\per \cm^2} are required to ensure that selected chromophores absorb one or a few  photons during their transit through the laser beam. 
Here, we demonstrate the realization of such optical gratings and discuss how the internal state evolution after the absorption process influences the evolution of the quantum wave that is associated with the molecular center-of-mass motion. \par

We specifically compare the four molecules shown in Fig.~\ref{fig1-molecules_processes}(a):
\textit{meso}-tetraphenylporphyrin (\ch{TPP}, $m=\SI{614.7}{u}$),
6,11-dihydroxy-5,12-naphthacenedione (\ch{DND}, and $m=\SI{290.3}{u}$), phthalocyanine (\ch{PcH2}, $m=\SI{514.5}{u}$) and a zinc-coordinated phthalocyanine where each isoindole unit is bound to an \textit{ortho}-nitro benzylic ether (NBE) group as a photocleavable tag (\ch{ZnPc-NBE4}, $m=\SI{1182.4}{u}$). \ch{TPP}, \ch{DND}, and \ch{PcH2} were obtained commercially (Sigma Aldrich/Merck) and used without further purification while \ch{ZnPc-NBE4} was synthesized by us based on a phthalocyanine core (see ESI). 
We use these different systems to explore the role of different molecular energy relaxation pathways some of which are indicated in the level scheme of Fig.~\ref{fig1-molecules_processes}(b).
They include internal conversion (IC), intersystem crossing (ISC), fluorescence, and the bond dissociation of a photocleavable tag. We discuss how these internal effects influence the de Broglie wave, i.e. the quantum evolution of the molecular center-of-mass motion, and how to observe it in experiments. \par

\section{Experimental setup}
The idea of the experiment is shown in Fig.~\ref{fig2-setup}.
All molecules are sublimated in a thermal source and the resulting beam is collimated to an angle below \SI{5}{\micro\radian}. Molecules of different velocity are spatially dispersed by their free-flight parabolas with a 20~µm high delimiter immediately behind the grating (not shown). 
This slit additionally ensures that all detected molecules have interacted with the light grating. The molecules propagate another \SI{0.7}{\meter} until they hit a thin quartz slide at the end of the vacuum chamber, where they are imaged using laser-induced fluorescence microscopy~\cite{juffmann_real-time_2012}, see ESI for details. \par

To realise the standing wave laser grating, \SI{5}{W} of laser radiation at $\lambda_{\rm L}=\SI{532}{nm}$ is frequency doubled in an external resonator (Sirah Wavetrain) to $\lambda_{\rm L}=\SI{266}{nm}$ with an output power of about \SI{1.2}{\watt}.
The DUV light is focused onto a dielectric mirror in high vacuum, with its surface aligned parallel to the molecular beam. 
To protect the laser from back-reflected light, and also to control the grating power, we employ an optical isolator, consisting of a $\lambda/2$ plate in front of a polarising beam splitter and a $\lambda/4$ plate behind it. The light in the optical grating is therefore circularly polarized. We track the power of the retro-reflected DUV beam and find that it is stable to within \SI{10}{\%} during a measurement.  
However, irradiating the mirror with light intensities beyond \SI{1}{\mega \watt \per \cm^2} at \SI{e-7}{\milli bar} leads to a slow degradation of the mirror surface.
To compensate for this, we shift the mirror parallel to the molecular beam in between measurements to expose a fresh spot to the laser. 
Given a grating period of $\lambda_{\rm L}/2=$\SI{133}{nm} and a laser waist of $12-15\,\mu$m~\cite{brand_fiber-based_2020}, the molecular beam divergence and its inclination to the mirror surface have to be smaller than $\SI{1}{mrad}$, to ensure that all molecules see a well-defined optical grating.  \par

While many aspects of matter-wave diffraction can be surprisingly well described using undergraduate level mathematics of waves~\cite{brand_single-_2021}, accounting for all experimental details and  molecular processes requires a full quantum description.
Our model accounts for the interaction between the molecules and the optical grating, the role of finite coherence and decoherence, the source collimation and velocity distribution and many internal relaxation pathways. The complete theory is based on propagating Wigner functions, as described in a separate paper~\cite{martinetz_probing_2024} and summarised in the ESI. Here, we focus on the conceptual discussion of the relevant processes. \par
As long as photon absorption can be neglected, the standing light wave acts as a pure phase grating: The interaction between the oscillating laser field and the dynamical polarisability $\alpha_\mathrm{266}$ of the molecule imposes a periodic dipole potential onto the molecular centre-of-mass motion, which modulates the phase of the molecular matter-wave along the $x$-axis:
\begin{equation}
    \Delta \phi (x)\propto  \frac{\alpha_\mathrm{266} P_{\rm L}}{\varepsilon_0 c w_y v_z}  \cos^2\left(\frac{2\pi x}{\lambda_{\rm L}}\right).
    \label{eq: dipole phase modulation}
\end{equation}
Here, $P_{\rm L}$ is the laser power, $w_y$ the vertical waist of the Gaussian laser beam, $v_z$ the forward molecular velocity, and $c$ the speed of light. The phase modulation  results in a discrete momentum transfer to the molecule in even multiples of the photon momentum  $\Delta p = \pm 2n\hbar k_{\rm L}$, where $ n\in\mathbb{N}$ and the photon wave number is $k_{\rm L}=2\pi/\lambda_{\rm L}$. This phase modulation of the matter-wave translates into a discrete spatial distribution of the molecular arrival probability density on the detector further downstream. This interaction is always present, since every molecule has a finite and sometimes even a large dynamical polarizability.\par
\begin{figure*}[th]
    \centering
    \includegraphics[width=0.9\linewidth]{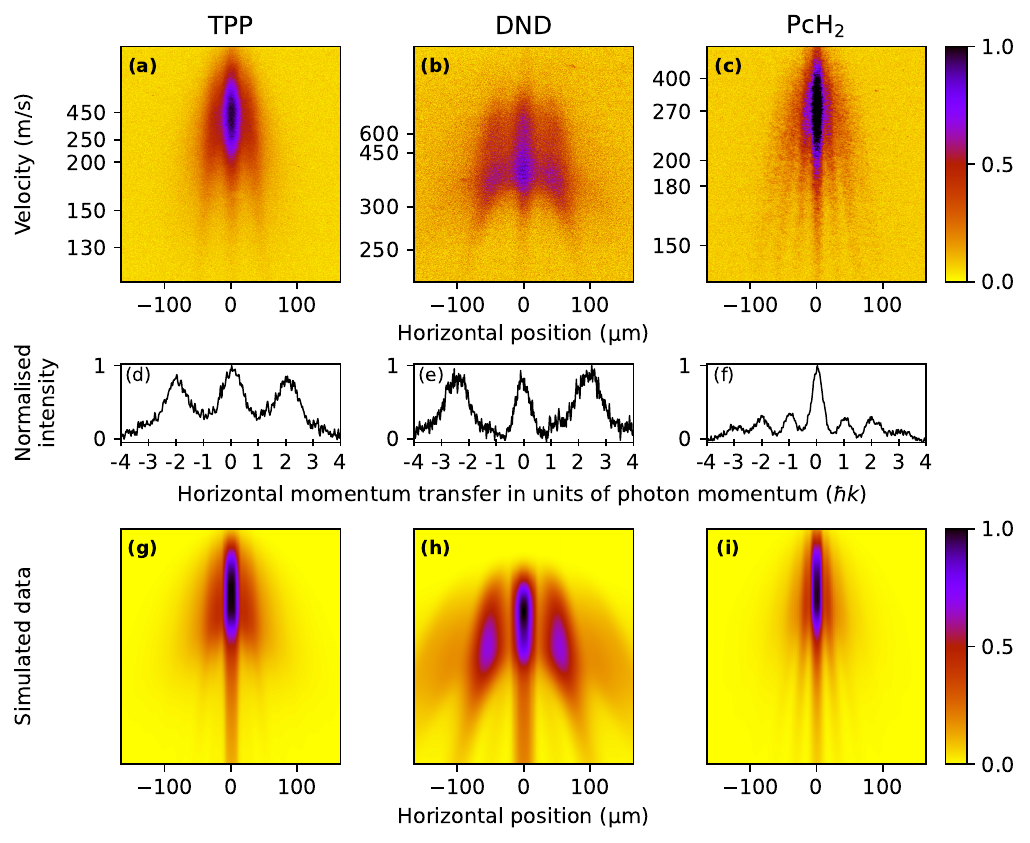}
    \caption{{\bf Top row:}  Fluorescence micrographs of the molecular interferograms: While the phase grating character dominates for TPP (a),  photon absorption gains importance for  phthalocyanine (c). {\bf Middle row:} Normalised traces for each of the fluorescence micrographs above, rescaled to the same momentum transfer and integrated over the lower two thirds of each image. {\bf Bottom row:} The numerical simulation shows good agreement with the experiment and allows corroborating the molecular ultraviolet polarizability and absorption cross section. }
    \label{fig:results-sim_comparison}
\end{figure*}
The description is more involved when the molecule can also absorb at least one photon from the laser grating. In this case, it receives an additional recoil of $\pm\hbar k_{\rm L}$ per  photon. This gives rise to additional peaks exactly half way in between the diffraction orders associated with the phase grating. Even though the absorption process is probabilistic and follows a Poisson distribution, it is coherent in the sense that one cannot, not even in principle, distinguish if the photon was absorbed while it was on the way towards the mirror or back. This is due to the long coherence length (here $50$\,m) of our DUV laser light~\cite{tomkovic_single_2011,cotter_coherence_2015}. 
At high intensities, absorption of $N$ photons can thus disperse the molecular momentum in integer multiples of the photon momentum, $\Delta p = n \hbar k_{\rm L}$ with $n=-N,\ldots,N$, and all branches of the molecular distribution associated with an even number of photons overlap at the detector position-synchronously with those affected by the phase grating alone, even though their internal state is different.\par
If an excited molecule decays non-radiatively, for instance by internal conversion (IC) or inter-system crossing (ICS) to a triplet state, the momentum transfer to the molecule is determined by the phase and absorption component alone. 
However, if spontaneous fluorescence is emitted near the grating, this adds again a momentum kick. Since the direction of spontaneously emitted photons is isotropically distributed, fluorescence would show up as a broadening of the diffraction peaks. Even multiple absorption-relaxation cycles are conceivable, given the range of absorption cross-sections and the laser intensities in our experiment.\par
Finally, the energy of a single or several photons may suffice to cleave the molecule. Our design and synthesis of ZnPc-NBE$_4$ was based on the idea that molecules should be selectively removed from the molecular beam upon photo-cleavage in the antinodes of the light grating and the fragments would be kicked to beyond the acceptance angle of the fluorescence detector (\SI{0.5}{\milli \radian}). 

\section{Results and Discussion}
We start our discussion with the diffraction pattern of TPP, as shown in Fig.~\ref{fig:results-sim_comparison}(a). 
It encompasses molecular velocities from \num{130} to beyond \SI{450}{\meter\per\second} which are dispersed on the detector because of their free fall in the gravitational field.
Based on our knowledge of the de Broglie wavelength $\lambda_{\rm dB}=h/(mv)\approx \SIrange[range-phrase=-]{4}{5}{\pico\meter}$ for the slower molecules and the grating period $d=\lambda_{\rm L}/2$, we can attribute diffraction to the effect of a dipole phase grating ($\Delta p = n 2\hbar k$), with only little background.  
To search for an effect of photo absorption and emission we show the results for DND  in Fig.~\ref{fig:results-sim_comparison}(b). The wider separation of the fringes is due to its smaller  molecular weight and larger de Broglie wavelength. And here, we observe well-defined peaks half-way in between the  diffraction orders associated with the phase grating. 
As all peaks have about the same width we see no indication for molecular fluorescence near the grating.\par

In solution, phthalocyanine is measured to have a good absorption cross section of $\sigma_{\rm 266} = \SI{1.1e-16}{\cm^2}$. We show its  diffraction pattern in Fig.~\ref{fig:results-sim_comparison}(c). Indeed, single-photon recoil shows up as clearly discernible peaks of the transverse momentum, at multiples of the photon momentum $\pm 1\hbar k_{\rm L}$. Again, the diffraction fringes are all equal, indicating that fluorescence plays no major role.
Because of their absorption properties, phthalocyanine derivatives are interesting candidates for photo-cleavage studies, when we decorate the core unit with four photo-reactive ortho-nitroso benzaldehyde group (NBE). Their photon absorption cross section in solution is even about a factor of two higher than that of the phthalocyanine core alone.
Earlier studies have already shown that a photo-reaction can release an ortho-nitroso benzaldehyde selectively even from a protein in the gas phase~\cite{schatti_neutralization_2019}.
Our matter-wave diffraction experiments with this molecule yield a diffraction image that resembles the result of unlabelled phthalocyanine. This invites two complementary interpretations which are discussed in detail in the ESI. The similarity of the \ch{PcH2} and \ch{ZnPc-NBE4} diffraction images can be explained within the frame work of matter-wave diffraction by the magnitude of the molecular de Broglie wavelengths and the expected photodepletion process. Another option is that the molecule may thermally decompose entirely in the source already.   
We find that here the thermal fragmentation precedes the optical dissociation, which demonstrates the high sensitivity of the NBE groups to the addition of internal energy. 
Since similar molecules are known to survive ultrafast laser evaporation in particular when injected into a cooling carrier gas or electrospray ionisation, photo-cleavage is still a promising basis for a deep ultraviolet beam splitter. The relevant effects and the theory apply as described above.  This insight opens a path to future explorations of peptide and protein interferometry.

\section{Conclusion}
We have shown that a deep ultra-violet standing light wave can act as a versatile beam splitter for organic molecules. This opens the door to the manipulation of novel particles and also to acquiring new information on photo-physical processes in molecules in the gas phase. 
Compared to typical spectroscopy methods the deactivation process is not encoded in the population of a final state but in the molecular center-of-mass motion, i.e. the  spatial diffraction pattern, where we can detect each molecule in principle with single molecules sensitivity~\cite{juffmann_real-time_2012}. While the signal-to-noise in our present setup is still too low for precision measurements, better sources will validate the concept in the future. 
The rich set of internal states of molecules will also allow us to explore a variety of additional photo-physical and photo-chemical effects for beam splitting and molecular analysis: For instance, when molecules are optically excited to long-lived triplet states, beam deflection in a magnetic field can be sensitively read out from interference patterns. Similarly, photo-isomerisation in the DUV grating will serve as a measurement-induced beam splitter when the detector at the end is sensitive and specific with regard to the molecular conformers.
We envisage that intense deep UV light gratings will become important building blocks for many all-optical matter-wave interferometers, designed to  explore molecular quantum optics in the regime of high mass and high complexity.

\section*{Author Contributions}
\noindent
Conceptualisation: MA, CB, KS, RF, MM. \par \noindent
Formal Analysis: LM, BS, KH, RF. \par \noindent 
Materials and synthesis: ADS, MM. \par \noindent
Funding acquisition: MA, KH, MM. \par \noindent
Investigation: KS, RF, LM, CB.\par \noindent
Methodology: all authors. \par \noindent
Software: LM, RF. \par \noindent
Supervision: MA, CB, KH, BS, MM. \par \noindent
Writing – original draft: MA, CB, KS, RF. \par \noindent

\section*{Conflicts of interest}
There are no conflicts to declare.

\section*{Data availability}
Data for this article and ESI, including raw, background-corrected and preprocessed diffraction images, as well as data files for simulated images, are available at zenodo repository under \href{https://doi.org/10.5281/zenodo.13124328}{https://doi.org/10.5281/zenodo.13124328}.
\section*{Acknowledgements}
We thank Y. Hua and V. Köhler for measuring the solution spectra of the photocleavable molecules and we thank D. Vörös, L. Gonz\'alez and A. Shayeghi for fruitful discussions. 
This research was funded in whole, or in part, by the Austrian Science Fund (FWF) [10.55776/DOC85] and [10.55776/P32543].
We acknowledge funding by the European Commission within project 860713, by the Gordon \& Betty Moore Foundation within project 10771, and by the German Aerospace Center (DLR) within project 50WM2264, with funds provided by the German Federal Ministry for Economic Affairs and Climate Action (BMWK). BAS acknowledges funding by the DFG--510794108 as well as by the Carl-Zeiss-Foundation through the project QPhoton. 
\providecommand*{\mcitethebibliography}{\thebibliography}
\csname @ifundefined\endcsname{endmcitethebibliography}
{\let\endmcitethebibliography\endthebibliography}{}

\end{document}


\title{Supplementary Information for\\Diffracting molecular matter-waves at deep-ultraviolet standing-light waves}

\author{Ksenija Simonovi\'c}
\email[]{ksenija.simonovic@univie.ac.at}
\affiliation{University of Vienna, Faculty of Physics, VDS, VCQ, Boltzmanngasse 5, A-1090 Vienna, Austria}

\author{Richard Ferstl}
\affiliation{University of Vienna, Faculty of Physics, VDS, VCQ, Boltzmanngasse 5, A-1090 Vienna, Austria}

\author{Alfredo Di Silvestro}
\affiliation{Department of Chemistry, University of Basel, St. Johannsring 19, 4056 Basel, Switzerland}

\author{Marcel Mayor}
\affiliation{Department of Chemistry, University of Basel, St. Johannsring 19, 4056 Basel, Switzerland}

\author{Lukas Martinetz}
\affiliation{University of Duisburg-Essen, Lotharstraße 1, 47048 Duisburg, Germany}

\author{Klaus Hornberger}
\affiliation{University of Duisburg-Essen, Lotharstraße 1, 47048 Duisburg, Germany}

\author{Benjamin A. Stickler}
\affiliation{Ulm University, Institute for Complex Quantum Systems and Center for Integrated Quantum Science and Technology, Albert-Einstein-Allee 11, 89069 Ulm, Germany}

\author{Christian Brand}
\affiliation{German Aerospace Center (DLR), Institute of Quantum Technologies, Wilhelm-Runge-Straße 10, 89081 Ulm, Germany}

\author{Markus Arndt}
\email[]{markus.arndt@univie.ac.at}
\affiliation{University of Vienna, Faculty of Physics, VDS, VCQ, Boltzmanngasse 5, A-1090 Vienna, Austria}%

\date{\today}

\maketitle

\section{Imaging of the diffraction patterns}
The interference patterns are collected on a quartz slide and illuminated by a homogeneous diffuse laser beam. The laser wavelength is chosen to match an excitation resonance of the molecules. \ch{TPP} is excited by \SI{421}{\nano\meter} light, \ch{DND} by \SI{266}{\nano\meter} light, and \ch{PcH2} as well as \ch{ZnPc-NBE4} by \SI{661}{\nano\meter} laser light. The fluorescence band pass filters are chosen to match the molecular spectra: \SIrange{630}{670}{\nano\meter} for \ch{TPP}, \SIrange{506}{594}{\nano\meter} for \ch{DND}, \SIrange{698.5}{723.5}{nm} for \ch{PcH2} and \SIrange{672}{712}{nm} for \ch{ZnPc-NBE4}. All fluorescence images were collected for 5 minutes, taking care to eliminate ambient light. We see no evidence of laser-induced fluorescence bleaching during this time except for \ch{DND}. Even in that case, a 5-minute exposition allows to image with good signal-to-noise. The fluorescence is collected using a 20$\times$ Zeiss plan neofluoar objective and imaged by a tube lens \textbf{($f=\SI{164}{\milli\meter}$)} onto an Andor iXON 3 EMCCD camera, cooled to \SI{-75}{\degreeCelsius}. 

\section{Data processing}
To correct the background signal in our raw images a dark CCD image (no laser) is subtracted by default in the recording software. Additionally we also subtract bright images taken before and after the molecules are deposited. This allows to reduce the contribution of stray light as well as contaminations from dust particles. Regions where obvious strong contaminations cannot be eliminated this way were manually removed from the dataset. We also eliminate intensity spikes by removing the lowest and highest $10^{-5}$-quantile of the data. Because of small variations in the ambient light  the background subtraction process can still leave inhomogeneities in the image background. They are reduced by subtracting a linear fit to the image that is gained from outside the region of the diffraction pattern. Finally a small rotation of the camera is corrected by rotating the images by \SI{0.4}{\degree} and the data is cropped to the region of interest where the diffraction pattern is located.

\section{Simulations}
\subsection{Overview}
\begin{figure}[ht]
    \centering
    \includegraphics[width=\linewidth]{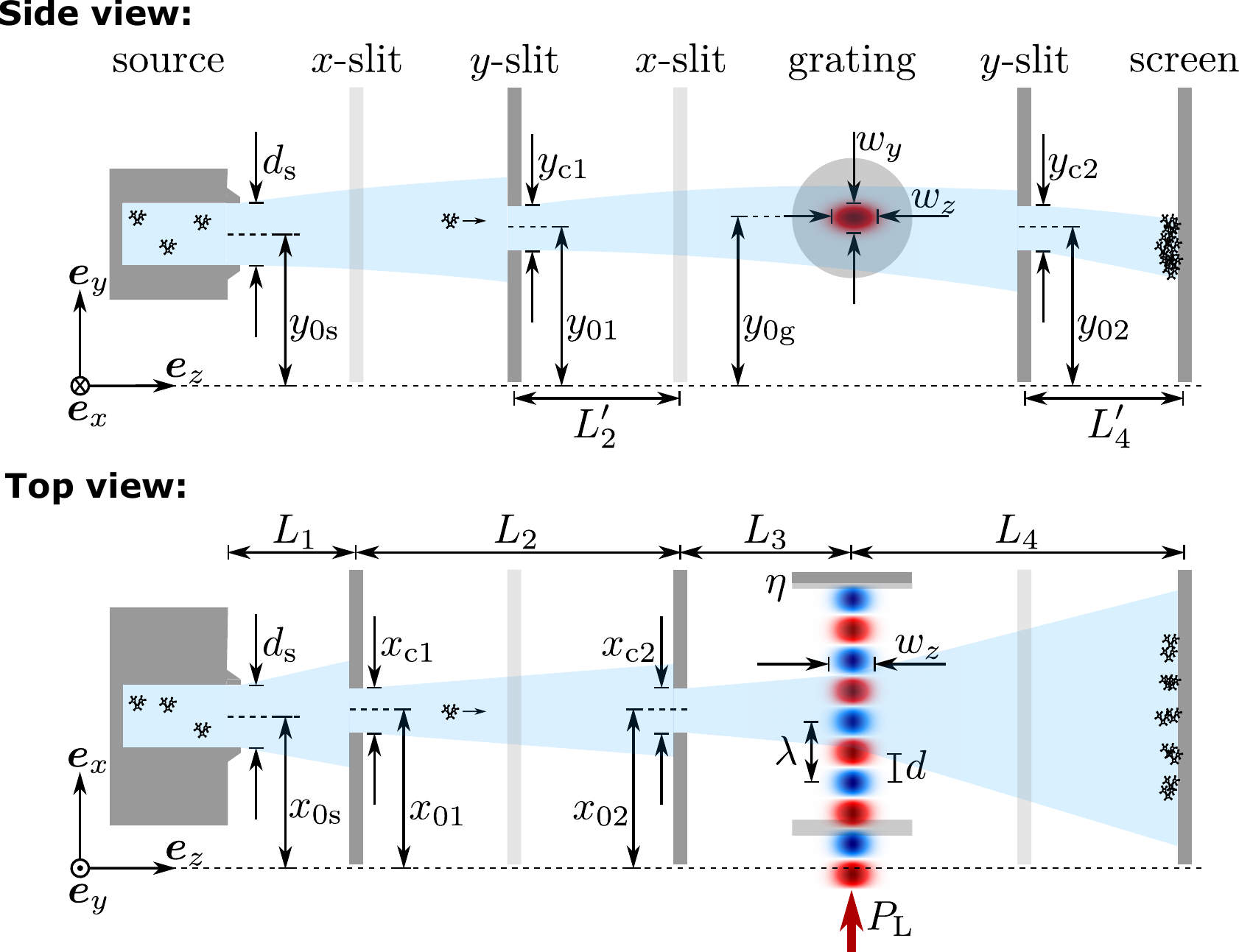}
    \caption{Schematic depiction of the experimental setup with all simulation-relevant parameters as listed in Tab.~\ref{tab:simulation_parameters} and the ESI text.}
    \label{fig:SI_setup-sketch}
\end{figure}
An in-depth derivation and explanation of the full quantum theory and all associated parameters are given in a separate manuscript~\cite{martinetz_probing_2024}. In Table~\ref{tab:simulation_parameters} we list the parameters that are required to  reproduce the simulated diffraction images of Fig.~3 in the main text. The interferometer setup with relevant parameters is also depicted in Fig.~\ref{fig:SI_setup-sketch}.

We first find the height of the velocity selection slit $y_{\rm 02}$ and the velocity shift $p_{\rm 0,z}/m$ for the expansion in the source to reproduce the horizontally integrated signal of the experiment. The height of the light grating $y_{\rm 0,g}$ and the optical properties $\alpha_{\rm r}$, $\sigma$ of the molecules are then optimised to find the best agreement with the experimental data. To find the optimum we  minimise the residual sum of squares over all pixels of the simulated and experimental image. The experimental data is vertically smoothed for this comparison.

For all cases we assume negligible fluorescence  $\phi_{\rm F}=0$, perfect reflectivity of the mirror $\eta = 1$ and a vanishing depletion probability. We further assume that internal conversion can be neglected in favour of intersystem crossing $\phi_{\rm IC}=0,\ \phi_{\rm ISC}=1$.
The gravitational acceleration is $g=-9.81$\, m/s$^2$, the rotation frequency of Earth at Vienna is $\omega \cdot \mathbf{e}_x=5.4\times10^{-5}{\si{\per\second}}$, and $\omega \cdot \mathbf{e}_y=-4.9\times10^{-5}{\si{\per\second}}$. The screen is discretised in squares of width $d_{\rm px}=\SI{0.33}{\micro m}$,  corresponding to the previously calibrated effective size of the pixels recorded by the CCD camera. The images are normalised to unity for the evaluation.

In the case of \ch{DND} the outlined optimisation procedure is impeded by a vertical bias of the signal most likely caused by inhomogeneous illumination of the diffraction image and the fact that no suitable background data is available. Here we manually set the required parameters to qualitatively reproduce the features of the experimentally observed diffraction image.

\begin{table*}[htb]
    \centering
    \begin{tabular}{|c|c|c|c|}
    \hline
    \textbf{Parameter} & \ch{TPP} & \ch{PcH2} & \ch{DND} \\ \hline
     Laser power $P_{\textrm L}$&\SI{0.92}{\watt}&\SI{0.96}{\watt}& \SI{1.0}{\watt}\\
	Laser waist $w_y$ & \SI{13}{\micro\meter}& \SI{16}{\micro\meter}& \SI{20}{\micro\meter}\\
	Molecule mass $m$ &\SI{614.74}{\atomicmassunit}& \SI{514.5}{\atomicmassunit} & \SI{290.3}{\atomicmassunit}\\
	DUV polarisability $\left|\alpha_{\textrm 266}\right|$&  $\SI{24}{\angstrom\cubed}\cdot 4\pi\varepsilon_0$ & $\SI{1.2}{\angstrom\cubed}\cdot 4\pi\varepsilon_0$& $\SI{35}{\angstrom\cubed}\cdot 4\pi\varepsilon_0$\\
	DUV absorption cross section $\sigma_{266}$&  \SI{3.4e-21}{\meter\squared} & \SI{8.5e-21}{\meter\squared}&  \SI{1e-21}{\meter\squared}\\
    Grating height $y_{\textrm 0g}$ &\SI{-12.7}{\micro \meter} &\SI{-5.1}{\micro \meter} & \SI{-43.0}{\micro \meter} \\ 
    Slit height $y_{\textrm 02}$ & \SI{-21.6}{\micro \meter} & \SI{-16.5}{\micro \meter}& \SI{-54.7}{\micro \meter}\\ 
    Source temperature $T$ & \SI{688}{\kelvin}& \SI{746}{\kelvin}& \SI{539}{\kelvin}\\ 
    Slit width $x_{\textrm c1}$ & \SI{3.5}{\micro\meter}& \SI{2.7}{\micro\meter}& \SI{5.0}{\micro\meter} \\
    Slit width $x_{\textrm c2}$ & \SI{1.7}{\micro\meter}& \SI{0.6}{\micro\meter}& \SI{3.0}{\micro\meter}\\
     Momentum shift $p_{0,z}/m$ & \SI{117}{\meter\per\second}&\SI{76.5}{\meter\per\second}&\SI{10.2}{\meter\per\second}\\
     \hline
Source size $d_{\textrm s}$ & \multicolumn{3}{c|}{\SI{200}{\micro\meter}}\\
Slit position $y_{01}$ & \multicolumn{3}{c|}{\SI{0}{\micro\meter}} \\
Slit width $y_{\textrm c1}$ &\multicolumn{3}{c|}{\SI{1}{\meter}} \\
Slit width $y_{\textrm c2}$ & \multicolumn{3}{c|}{\SI{20}{\micro\meter}}\\
        Distance $L_1$ & \multicolumn{3}{c|}{\SI{0.52}{\meter}}\\
        Distance $L_2$ & \multicolumn{3}{c|}{\SI{0.3}{\meter}}\\
        Distance $L_2'$ &\multicolumn{3}{c|}{\SI{0.02}{\meter}}\\
        Distance $L_3$ & \multicolumn{3}{c|}{\SI{0.08}{\meter}}\\
        Distance $L_4$ & \multicolumn{3}{c|}{\SI{0.69}{\meter}}\\
        Distance $L_4'$ &\multicolumn{3}{c|}{\SI{0.605}{\meter}}\\
	Grating period $d$ &\multicolumn{3}{c|}{\SI{133}{\nano\meter}}\\
    \hline
    \end{tabular}
    \caption{Parameters used to simulate the diffraction images shows in Fig.~3 of the main text.}
    \label{tab:simulation_parameters}
\end{table*}

\subsection{Bounds on accuracy and precision}
\begin{figure}[h]
    \centering
    \includegraphics[width=\linewidth]{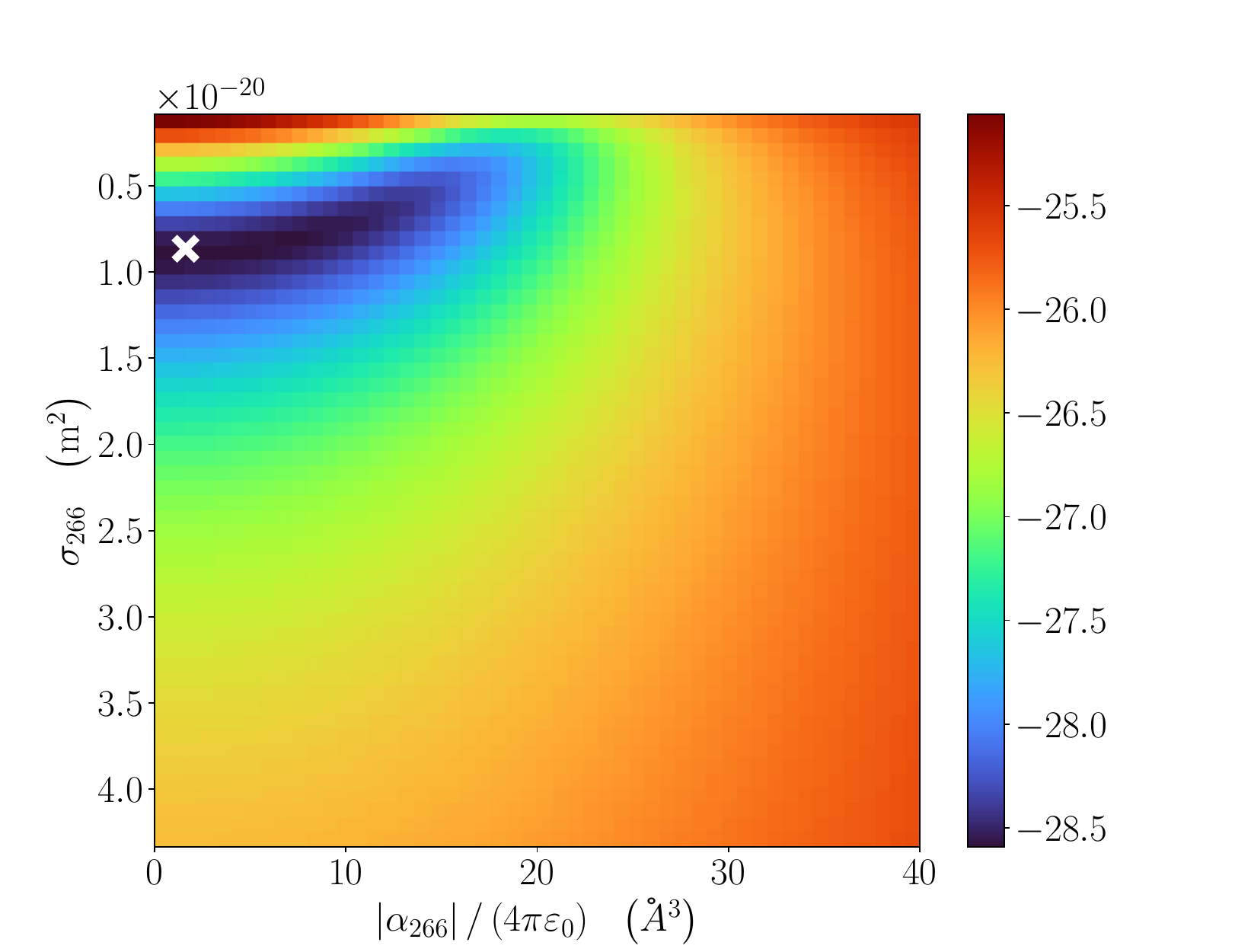}
    \caption{Natural logarithm of the residual sum of squares as a function of the parameters $\left|\alpha_{\rm 266}\right|$ and $\sigma_{\rm 266}$ computed for the simulated and experimental data of \ch{PcH2}. Lower values indicate a better agreement between the two datasets. The minimum is marked by a white cross.}
    \label{fig:pch2_scan}
\end{figure}

The optimisation procedure described above allows us to reproduce our experimental data and to assign the dominant diffraction processes. 
In Fig.~\ref{fig:pch2_scan} we show a scan of the possible deep ultraviolet molecular polarizabilities and absorption cross sections, $\alpha_{\rm r}$--$\sigma$, that fit the experiment.
The heat map shows the natural logarithm of the residual sum of squares computed for the simulation and experimental data. The agreement between experiment and theory allows to estimate  an order of magnitude for the DUV polarisabilities and cross sections. Note that the diffraction patterns are insensitive to the sign of the polarizability in this setup.
Complementary measurements shall be realised in a revised setup to extract these optical properties with improved accuracy and precision.

\section{Synthesis and characterisation of \ch{ZnPc-NBE4}}
\textbf{General Procedures:} All commercially available chemicals were used without further purification. Dry solvents were used as crown cap and purchased from \textsc{Acros Organics} and \textsc{Sigma-Aldrich}. NMR solvents were obtained from \textsc{Cambridge Isotope Laboratories, Inc.} (Andover, MA, USA) or \textsc{Sigma-Aldrich}. All NMR experiments were performed on \textsc{Bruker Avance III} or \textsc{III HD}, two- or four-channel NMR spectrometers operating at \num{400.13}, \num{500.13} or \SI{600.27}{MHz} proton frequency. The instruments were equipped with direct observe BBFO, indirect BBI or cryogenic four-channel QCI (\ch{H}/\ch{C}/\ch{N}/\ch{F}) \SI{5}{mm} probes, all with self-shielded z-gradient. The experiments were performed at \SI{298}{K} or \SI{295}{K}. All chemical shifts ($\delta$) are reported in parts per million (ppm) relative to the used solvent and coupling constants ($J$) are given in Hertz (\unit{Hz}). The multiplicities are written as: s = singlet, d = doublet, t = triplet, dd = doublet of doublet, m = multiplet. Flash column chromatography (FCC) was performed with SiliaFlash® P60 from SILICYCLE with a particle size of \SIrange{40}{63}{\micro m} (230-400 mesh), and for TLC silica gel 60 F254 glass plates with a thickness of \SI{0.25}{mm} from \textsc{Merck} were used. The detection was carried out with a UV-lamp at \num{254} or \SI{366}{nm}. UV/VIS absorption spectra were recorded on a \textsc{Jasco} V-770 Spectrophotometer with a \SI{1}{cm} quartz glass cuvette. High-resolution mass spectra (HRMS) were measured with a \textsc{Bruker Maxis 4G} ESI-TOF instrument or a \textsc{Bruker solariX} spectrometer with a MALDI source.

\textbf{Synthetic steps to the target structure:} As displayed in Scheme~\ref{sc:synthesis_esi}, the target structure \ch{ZnPc-NBE4} was assembled in two synthetic steps, followed by purification via sublimation. As the first step, the condensation between (2-nitrophenyl)methanol and 4-hydroxyphthalonitrile provided the phthalonitrile precursor (4-((2-nitrobenzyl)oxy)phthalonitrile) exposing the photo-cleavable nitro benzyl ether subunit. Its subsequent cyclotetramerization in the presence of zinc acetate provided the target structure as mixture of regioisomers. To our delight, sublimation of the crude mixture provided the highly symmetric regioisomer \textbf{\ch{ZnPc-NBE4}} which was used in the experiments reported here.  

\textbf{Synthesis of 4-((2-nitrobenzyl)oxy)phthalonitrile:} An oven dried \SI{100}{\milli \liter} round-bottomed flask under argon was charged with (2-nitrophenyl)methanol (\SI{1831}{mg}, \SI{11.6}{\milli \mol}, \SI{1.0}{eq}.),  4-hydroxyphthalonitrile (\SI{2006}{mg}, \SI{13.9}{\milli\mol}, \SI{1.2}{eq}.), triphenylphosphine (\SI{4564}{mg}, \SI{17.4}{\milli \mol}, \SI{1.5}{eq}.) and dry tetrahydrofuran (\ch{THF}, \SI{50}{\milli \liter}). The reaction mixture was cooled to \SI{0}{\celsius} and diisopropyl azodicarboxylate (\ch{DIAD}, \SI{3.43}{mL}, \SI{17.4}{\milli \mol}, \SI{1.5}{eq}.) was added drop wise. The reaction was allowed to warm up to room temperature and stirred for 12 hours. After TLC confirmed full conversion of the starting materials the solvent was removed under reduced pressure and the remaining residue was purified by column chromatography (ethyl acetate/cyclohexane = 1/2) to yield the desired product as an off-white solid (\SI{2000}{mg}, \SI{11.6}{\milli \mol}, \SI{61}{\%}). \par   
\textbf{Analytical data for 4-((2-nitrobenzyl)oxy) phthalonitrile}:
$^1$H NMR (\SI{400}{MHz}, \ch{DMSO-d6}, \SI{22}{\celsius}) $\delta$ = 8.18 (dd, J = 8.2, \SI{1.3}{Hz}, 1H), 8.08 (d, J = \SI{8.8}{Hz}, 1H), 7.91 (d, J = \SI{2.7}{Hz}, 1H), 7.81 (td, J = 7.6, \SI{1.3}{Hz}, 1H), 7.74 (dd, J = 7.8, \SI{1.5}{Hz}, 1H), 7.66 (ddd, J = 8.7, 7.4, \SI{1.5}{Hz}, 1H), 7.57 (dd, J = 8.9, \SI{2.6}{Hz}, 1H), 5.66 (s, 2H).
$^{13}$C NMR (101 MHz, DMSO-d$_6$, 22°C) $\delta$ = 161.18, 147.41, 135.92, 134.25, 131.07, 129.57, 129.21, 125.08, 120.60, 120.44, 116.37, 116.10, 115.63, 106.69, \SI{67.51}{ppm}. 
HRMS (ESI-ToF, MeOH, positive mode): calc. for \ch{[C60H36N12O12Zn]+} 302.0536; found 302.0537. \par
\textbf{Synthesis of \ch{ZnPc-NBE4}:} An oven dried \SI{50}{mL} round-bottomed flask under argon was charged with 4-((2-nitrobenzyl)oxy)phthalonitrile (\SI{2000}{mg}, \SI{7.16}{\milli\mol}, \SI{1.0}{eq}.), urea (\SI{1247}{mg}, \SI{20.8}{\milli\mol}, \SI{2.9}{eq}.), zinc acetate (\SI{328}{mg}, \SI{1.79}{\milli\mol}, \SI{0.28}{eq}.), ammonium molybdate (\SI{28}{mg}, \SI{0.14}{\milli\mol}, \SI{0.02}{eq}.) and were heated in nitrobenzene (\SI{30}{mL}) at \SI{160}{\celsius} for \SI{12}{h}. After cooling, the reaction mixture was treated with water and the green dark product that precipitated was filtered off, successively washed with water, ethyl acetate and ethanol. After evaporation of the solvent under reduced pressure, the crude was purified by column chromatography (acetone/pyridine 5/1 and later 1/1) to obtain the mixture of regioisomers as a blue solid (\SI{460}{mg}, \SI{1.79}{mL}, \SI{21}{\%}). Sublimation provided the pure regioisomer in yields below \SI{5}{\%}.\par
\textbf{Analytical data for \ch{ZnPc-NBE4}}: 
$^1$H NMR (\SI{500}{MHz}, \ch{DMSO-d6}, \SI{22}{\celsius}) $\delta$ = 9.25 (d, J = \SI{6.8}{Hz}, 4H), 8.91 (d, J = \SI{20}{Hz}, 4H), 8.29 (t, J = \SI{7.4}{Hz}, 4H), 8.19 (d, J = \SI{7.5}{Hz}, 4H), 7.79 (t, J = \SI{7.6}{Hz}, 4H), 7.88 (s, 4H), 7.77 (t, J = \SI{7.8}{Hz}, 4H), 6.05 (s, 8H). HRMS (MALDI, DCM/DCTB Mix 1:10, positive mode): calc. for \ch{[C60H36N12O12Zn]+} 1180.1862; found 1180.1858.\par
\begin{scheme}[ht]
   \centering
    \includegraphics[width=0.75\linewidth]{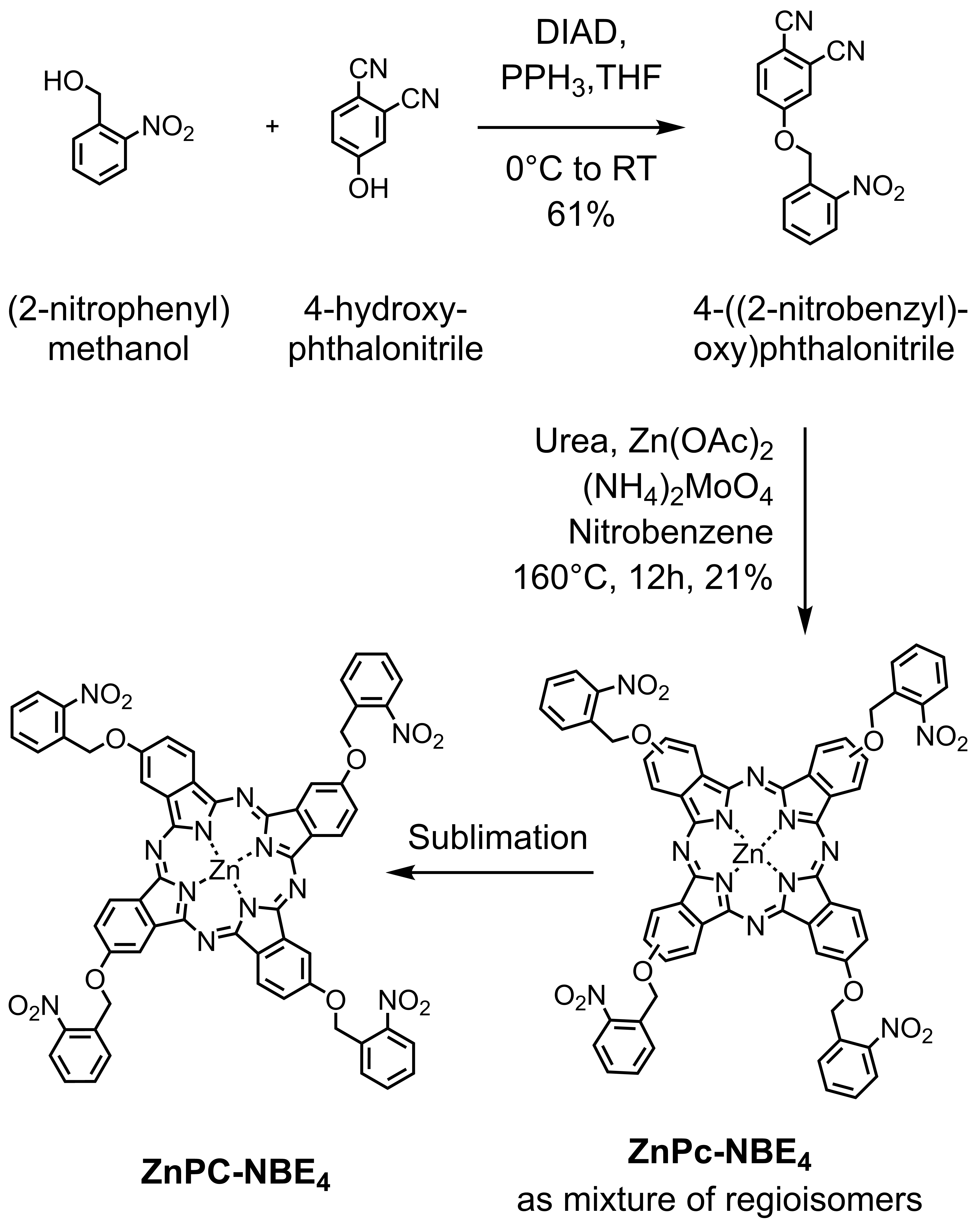}
    \caption{Synthesis scheme for the compound \textbf{\ch{ZnPc-NBE4}}.}
    \label{sc:synthesis_esi}
\end{scheme}
\section{Measurements with \ch{ZnPc-NBE4}}
\begin{figure*}[t]
    \centering
    \includegraphics[width=\linewidth]{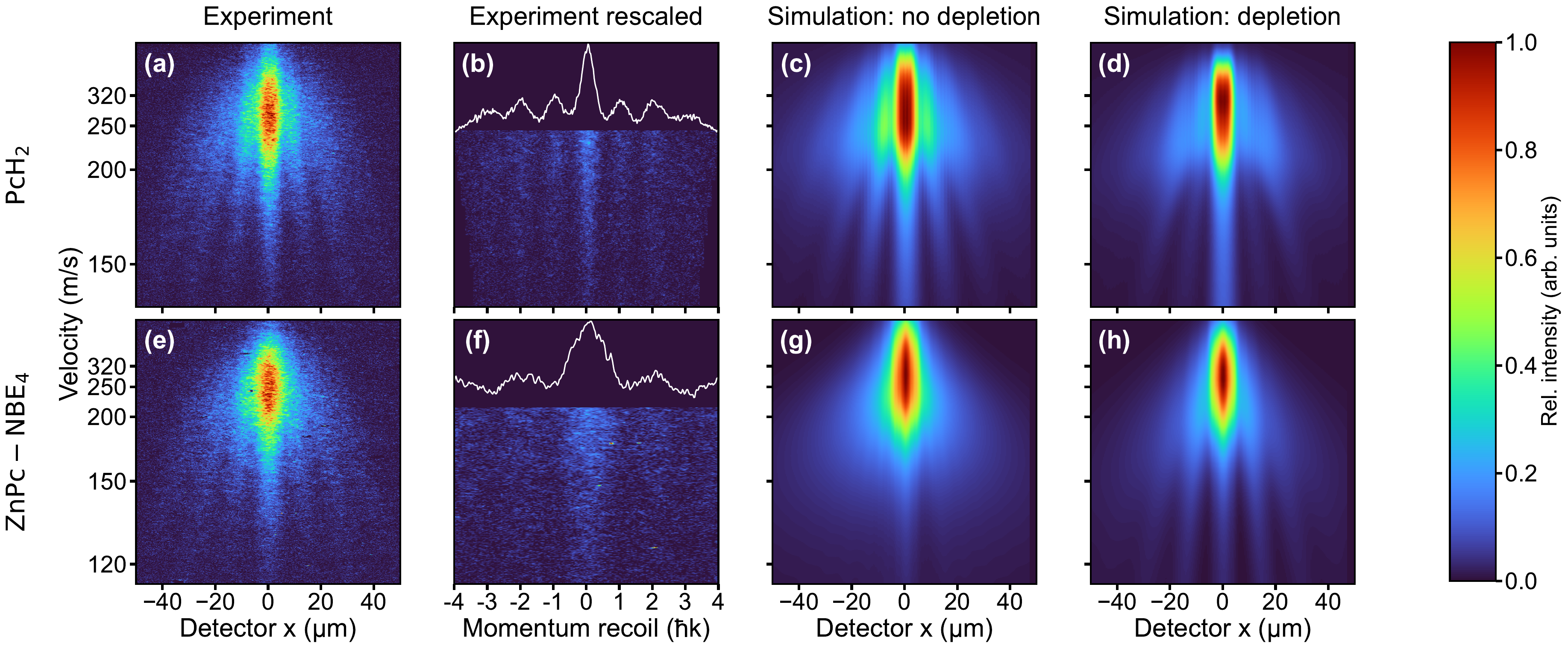}
    \caption{Comparing the experimental and simulated diffraction patterns of \ce{PcH2} (top row) and \ce{ZnPc-NBE4} (bottom row). The measured patterns (\textbf{(a)}, \textbf{(e)} are nearly identical, suggesting cleavage of \ce{ZnPc-NBE4}. This is further supported by simulation considering effects of depletion for \ce{ZnPc-NBE4} (\textbf{(g)}, \textbf{(h)}), where only the latter (assuming the efficient depletion) reproduces the measured data. Note that the opposite is true for \ch{PcH2}.}
    \label{fig:phthalo-cleavables-comp}
\end{figure*}
Interestingly, we find that the the diffraction pattern for \ch{ZnPc-NBE4} is almost identical to that of \ch{PcH2}, as shown in Fig.~\ref{fig:phthalo-cleavables-comp}. This invites two complementary interpretations: First, numerical simulations with and without photodissociation of \ch{ZnPc-NBE4} show that the observed fringe pattern can be explained under the assumption that photocleavage is present and efficient (Fig.~\ref{fig:phthalo-cleavables-comp}~(g-h)). This is true independent of whether a single, two, three, or all functional groups are split off, if only the intact parent molecules make it to the detector and all fragments are kicked to beyond the detector acceptance angle. The diffraction pattern would look similar to that of \ch{PcH2} because effective cleavage via single photon absorption would remove the absorption fringes. The $n \hbar k$ peaks of \ch{PcH2} (Fig.~\ref{fig:phthalo-cleavables-comp}~(a-d)) would therefore be practically co-located with the $2n\hbar k$ peaks of \ch{ZnPc-NBE4}. The second interpretation, however, is also attractive: at a temperature of \SI{400}{\celsius} all four \ch{NBE} groups as well as the coordinated \ch{Zn} atom may already be detached in the thermal source. In this case, the diffraction patterns look identical because the molecules are identical. \par
\begin{figure*}[ht]
    \centering
    \includegraphics[width=0.8\linewidth]{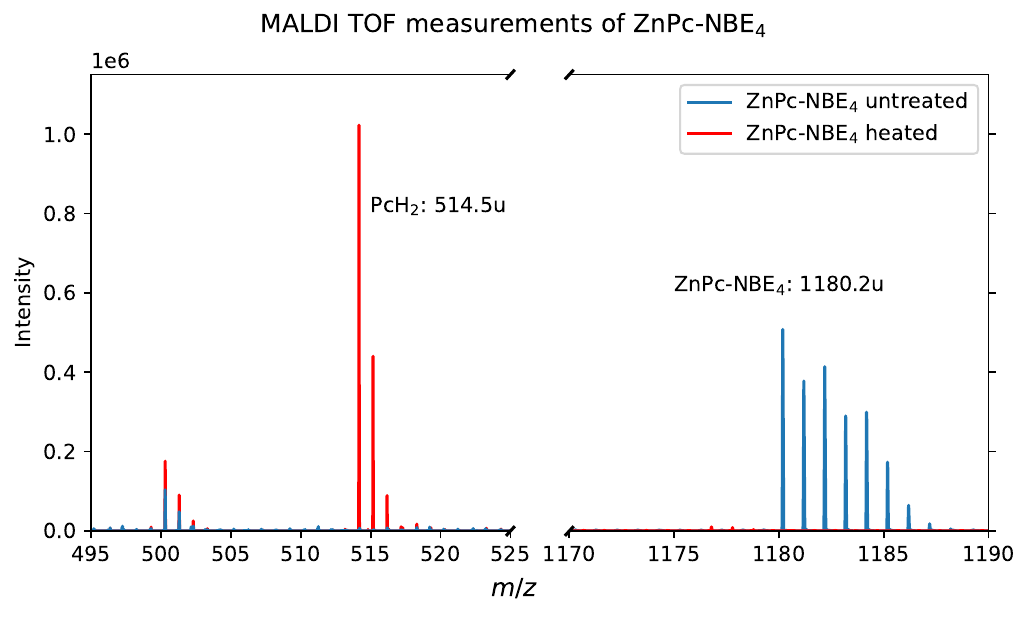}
    \caption{MALDI-TOF measurements of untreated \ch{ZnPc-NBE4} (blue) and after oven sublimation at temperatures needed to create a molecular beam. While the untreated sample shows a strong peak around \SI{1180}{u} as expected, the heated one only shows a peak at \SI{514.5}{u}. This is the mass of the metal-free phthalocyanine core, indicating that the sample undergoes nearly complete thermal decomposition in the source already.}
    \label{fig:SI-MALDI}
\end{figure*}

To distinguish between these two possibilities one can envisage two tests, one based on matter-wave arguments and one using mass spectrometry. Even though the peaks are co-located, the intensity distribution of the interference fringes should depend on the optical polarizability of the arriving molecules - which is substantially bigger for \ch{ZnPc-NBE4} than for \ch{PcH2}.
However, since DUV polarisabilities in the gas phase are not available from independent measurements, this interesting route remains closed for now. 
Re-collecting the emitted molecules on a glass slide behind the oven and post-analysing them in MALDI-MS shows that thermal decomposition is almost complete - encompassing all NBE subgroups down to the bare phthalocyanine core as shown in Fig~\ref{fig:SI-MALDI}. This underlines the importance of developing nondestructive, bright sources for molecular beams that would allow conducting these and similar experiments with thermally fragile systems.   

\providecommand*{\mcitethebibliography}{\thebibliography}
\csname @ifundefined\endcsname{endmcitethebibliography}
{\let\endmcitethebibliography\endthebibliography}{}